\documentclass[conference]{IEEEtran}
\IEEEoverridecommandlockouts
\usepackage{amsmath,amssymb,amsfonts}
\usepackage{graphicx}
\usepackage{textcomp}
\usepackage{xcolor}
\usepackage[nolist,nohyperlinks]{acronym}
\usepackage{lipsum}
\usepackage{booktabs} 

\usepackage{hyperref}

\usepackage{algorithm2e} 

\usepackage{float}
\usepackage{nicefrac}
\usepackage{thmtools}
\usepackage{enumerate}
\usepackage[caption=false,font=footnotesize]{subfig}

\def\BibTeX{{\rm B\kern-.05em{\sc i\kern-.025em b}\kern-.08em
    T\kern-.7667em\lower.7ex\hbox{E}\kern-.125emX}}

\def\mathlette#1#2{{\mathchoice{\mbox{#1$\displaystyle #2$}}%
                           {\mbox{#1$\textstyle #2$}}%
                           {\mbox{#1$\scriptstyle #2$}}%
                           {\mbox{#1$\scriptscriptstyle #2$}}}} 
\renewcommand{\Vec}[1]{\mathlette{\boldmath}{#1}}

\begin{document}

%\title{Mobile User Equipment Tracking with mmWave RIS for Industrial Automation}
%\title{RIS Coverage for Mobile User Equipment in Industrial Automation: Analysis and Experiment}  %Mentioning mmWave somewehere in title?
\title{Reconfigurable Intelligent Surface for Industrial Automation: mmWave Propagation Measurement, Simulation, and Control Algorithm Requirements}
\author{Hamed Radpour$^\ast$, Markus Hofer$^\ast$, David Löschenbrand$^\ast$, Lukas Walter Mayer$^\dagger$, Andreas Hofmann$^\dagger$, \\ Martin Schiefer$^\dagger$ and Thomas Zemen$^\ast$\\
	$^\ast$AIT Austrian Institute of Technology, Vienna, Austria\\
    $^\dagger$Siemens Aktiengesellschaft Oesterreich, Vienna, Austria\\
	Email: hamed.radpour@ait.ac.at}
	
%\makecoverpage
%\cleardoublepage
	
\maketitle
\begin{acronym} 
\setlength{\itemsep}{-0.63\parsep}
\acro{3GPP}{3rd Generation Partnership Project}
%\acro{AOA}{angle of arrival}
%\acro{AOD}{angle of departure}
\acro{ACF}{auto correlation function}
%\acro{ASIC}{application-specific integrated circuit}
\acro{ADC}{analog-to-digital converter}
%\acro{AWGN}{additive white Gaussian noise}
\acro{BEM}{basis-expansion model}
%\acro{BLER}{block error rate}
%\acro{BER}{bit-error-rate}
%\acro{CA}{complex addition}
\acro{CDF}{cumulative distribution function}
\acro{CE}{complex exponential}
\acro{CIR}{channel impulse response}
%\acro{CDMA}{code division multiple access}
%\acro{CM}{complex multiplication}
%\acro{COTS}{common-off-the-shelf}
\acro{CP}[CP]{cyclic prefix}
%\acro{CSM}{complex scalar multiplication}
\acro{CTF}{channel transfer function}
\acro{D2D}{device-to-device}
\acro{DAC}{digital-to-analog converter}
\acro{DC}{direct current}
%\acro{DDC}{direct down conversion}
%\acro{DMA}{direct memory access}
\acro{DOD}{direction of departure}
\acro{DOA}{direction of arrival}
%\acro{DUC}{direct up conversion}
\acro{DPS}{discrete prolate spheroidal}
\acro{DPSWF}{discrete prolate spheroidal wave function}
\acro{DSD}{Doppler spectral density}
\acro{DSP}{digital signal processor}
\acro{DR}{dynamic range}
%\acro{eNodeB}{base station}
%\acro{eMBB}{enhanced mobile broadband} %used
\acro{ETSI}{European Telecommunications Standards Institute}
%\acro{FER}{frame error rate}
\acro{FIFO}{first-input first-output}
\acro{FFT}{fast Fourier transform}
%\acro{FLOPS}{floating point operations per second}
\acro{FP}{fixed-point}
\acro{FPGA}{field programmable gate array}
%\acro{FPS}{frames per second}
%\acro{GCE}{geometry-based channel emulator}
\acro{GSCM}{geometry-based stochastic channel model}
%\acro{GCM}{geometry-based channel model}
\acro{GSM}{global system for mobile communications}
\acro{GPS}{global positioning system}
%\acro{HiL}{hardware-in-the-loop}
\acro{HPBW}{half power beam width}
\acro{ICI}[ICI]{inter-carrier interference}
\acro{IDFT}{inverse discrete Fourier transform}
\acro{IF}{intermediate frequency}
\acro{IFFT}{inverse fast Fourier transform}
\acro{ISI}{inter-symbol interference}
\acro{ITS}{intelligent transportation system}
%\acro{mMTC}{massive machine type communications} %used
%\acro{MAC}{multiply-accumulate}
\acro{MEC}{mobile edge computing}
\acro{MSE}{mean square error}
\acro{LLR}{log-likelihood ratio}
\acro{LO}{local oscillator}
\acro{LOS}{line-of-sight}
\acro{LMMSE}{linear minimum mean squared error}
\acro{LNA}{low noise amplifier}
\acro{LSF}{local scattering function}
\acro{LTE}{long term evolution}
%\acro{LTE-A}{Long Term Evolution Advanced}
\acro{LUT}{look-up table}
\acro{LTV}{linear time-variant }
\acro{MIMO}{multiple-input multiple-output}
\acro{MPC}{multi-path component}
\acro{MC}{Monte Carlo}
\acro{NI}{National Instruments}
\acro{LoS}{line-of-sight}
\acro{NLOS}{non-line of sight}
\acro{OFDM}{orthogonal frequency division multiplexing}
\acro{OTA}{over-the-air}
\acro{PA}{power amplifier}
\acro{PC}{personal computer}
\acro{PDP}{power delay profile}
%\acro{PDF}{probability density function}
\acro{PER}{packet error rate}
\acro{PPS}{pulse per second}
%\acro{P2P}{peer-to-peer}
\acro{QAM}{quadrature ampltiude modulation}
\acro{QPSK}{quadrature phase shift keying}

%\acro{UE}{user equipment}
\acro{RB}{resource block}
\acro{RBP}{resource block pair}
\acro{RIS}{reflective intelligent surface}
\acro{RF}{radio frequency}
\acro{RMS}{root mean square}
\acro{RSSI}{receive signal strength indicator}
\acro{RT}{ray tracing}
\acro{RX}{receiver}
\acro{SCME}{spatial channel model extended}
\acro{SDR}{software defined radio}
\acro{SISO}{single-input single-output}
\acro{SoCE}{sum of complex exponentials}
\acro{SNR}{signal-to-noise ratio}
\acro{SUT}{system-under test}
\acro{SSD}{soft sphere decoder}
\acro{TBWP}{time-bandwidth product}
\acro{TDL}{tap delay line}
%\acro{TTI}{transmission time intervall}
\acro{TX}{transmitter}
%\acro{UAV}{unmanned ariel vehicle}
\acro{UMTS}{universal mobile telecommunications systems}
\acro{UDP}{user datagram protocol}
\acro{URLLC}{ultra-reliable and low latency communication} %yes
\acro{US}{uncorrelated-scattering}
\acro{USRP}{universal software radio peripheral}
\acro{VNA}{vector network analyzer}
\acro{ViL}{vehicle-in-the-loop}
\acro{V2I}{vehicle-to-infrastructure}
\acro{V2V}{vehicle-to-vehicle}
\acro{V2X}{vehicle-to-everything}
\acro{VST}{vector signal transceiver}
\acro{VTD}{Virtual Test Drive}
\acro{WF}{Wiener filter}
\acro{WSS}{wide-sense-stationary}
\acro{WSSUS}{wide-sense-stationary uncorrelated-scattering}
\end{acronym}

 \begin{abstract}
Reconfigurable intelligent surfaces (RISs) can provide a reliable and low-latency millimeter wave (mmWave) communication link in cases of a blocked line-of-sight (LoS) between the base station (BS) and the user equipment (UE). In such cases, the RIS mounted on a wall or ceiling can act as a bypass for the radio communication link. We present an active RIS with 127 patch antenna elements arranged in a hexagonal grid. The RIS operates at a center frequency of 23.8\,GHz and each RIS element uses an orthogonal polarization transformation to enable amplification using a field-effect transistor (FET). The source and drain voltages of each FET are controlled using two bits. We consider that the RIS control unit is aware of the UE coordinates within the measurement area, relevant to the industrial control scenarios. We measure the received power on a 2D grid of 60\,cm$\times$90\,cm using an $xy$-positioning table, with the RIS working in reflective and active mode. The results demonstrate the ability of the RIS to effectively focus the radio signal on the desired target points. We characterize the half-power beam width in azimuth and radial directions with respect to the RIS position, enabling us to obtain a practical RIS configuration update criterion for a mobile UE. These results clearly show that RISs are prominent solutions for enabling reliable wireless communication in indoor industrial scenarios.
\end{abstract}


\section{Introduction}
In this paper, we are specifically interested in using a reconfigurable intelligent surface (RIS) in an indoor automation and control scenario with high-reliability and low-latency requirements for a mmWave communication link. The RIS shall provide a bypass for the radio signal in case the line-of-sight (LoS) is blocked between the base station (BS) and the user equipment (UE).

The authors of \cite{zhi2022active,Radpour24} show that an active RIS outperforms a passive RIS under the same power budget. Hence, active RIS elements can either increase the signal-to-noise ratio (SNR) at the UE when the number of RIS elements is kept constant, or the number of RIS elements can be reduced for a constant SNR target, helping to reduce the cost and size of the RIS. Moreover, a RIS with a smaller number of elements can quickly update its element configuration to track a mobile UE. This feature is particularly advantageous for time-sensitive automation and control applications, which are the main focus of this research paper.  

In \cite{Radpour24} an active RIS for the mmWave frequency band is presented using a polarization transformation to enhance the isolation between incoming and reflected waves. Numerical simulations and measurements for the received power show a good match for the radiation pattern in an anechoic chamber. However, validation and testing of RISs in indoor non-LoS (NLoS) scenarios with mobile user equipment (UE) is necessary for industrial automation to ensure a reliable link between the base station (BS) and UE.

The authors of \cite{liu2023reconfigurable} investigate a RIS-assisted indoor scenario with moving robots. For this work, they assumed to have either a large number of small RISs or a small number of large RISs. In \cite{bian2023reconfigurable}, a mobile target tracking scenario is considered by estimating motion parameters. However, these two studies are theoretical and need to be verified with measurements in practical environments. The authors in \cite{kayraklik2023indoor} studied a RIS grouping approach using an iterative algorithm and performed measurement for sub-6 GHz.

\subsection*{Scientific Contributions}
\begin{itemize}
	\item We present an enhanced version of the RIS described in \cite{Radpour24} containing $127$ patch antenna elements for the mmWave band with optimized reflection coefficients. We evaluate the ability of the RIS to focus the radio signal on the mobile UE location, analyzing a 2D measurement area of 60\,cm$\times$90\,cm using a $xy$-positioning table. We compare the measurement results with a numerical simulation, achieving an excellent qualitative and quantitative match. 
  \item The obtained received power pattern reveals how often we need to update the RIS configuration to maintain the received power at the mobile UE location above a certain threshold. We compare the results of the RIS in active and reflective mode.
 \item Results show that the proposed RIS can effectively focus the radio signal on a mobile UE in a NLoS scenario.
\end{itemize}

\section{RIS System Model}
\begin{figure}
	\centering
	\includegraphics[width=\columnwidth]{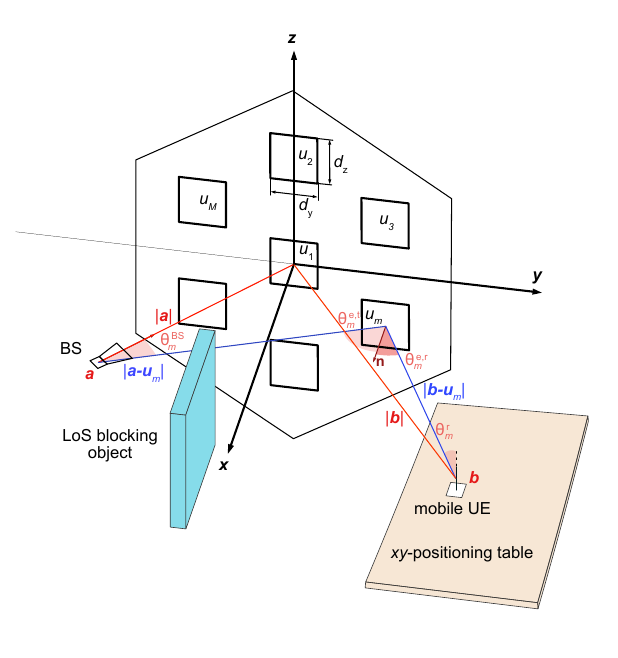}
 \vskip-18pt
	\caption{RIS coordinate system for a hexagonal RIS element placement in the $yz$-plane. The BS horn antenna radiates from position $\Vec{a}$ towards the center of the RIS at $\Vec{0}=(0,0,0)$ over a distance of $|\Vec{a}|$. The UE monopole antenna at position $\Vec{b}$ is within a distance of $|\Vec{b}|$ moving on a $xy$-positioning table. The LoS is blocked between the BS and the UE. The picture is not to scale to improve clarity.} 
	\label{fig:RIScoordinates}
 \vskip-18pt
\end{figure}

For our system model, we consider that there is no direct link between BS and UE and there is a line-of-sight (LoS) channel for the BS-RIS and the RIS-UE link. The BS is located at $\Vec{a}=(a_\text{x},a_\text{y},a_\text{z})$ in Cartesian coordinates. We also use spherical coordinates $\tilde{\Vec{a}}=(a_r, a_\varphi, a_\theta)$ depending on the context. We measure azimuth $a_\varphi$ between the x-axis and the projection of the vector $\tilde{\Vec{a}}$ onto the $xy$-plane and elevation $a_\theta$ between the projection in the $xy$-plane and the vector itself; see Fig. \ref{fig:RIScoordinates}. 
%\vskip-10pt
The UE is located at $\Vec{b}$ and the $M$ RIS element center points at $\Vec{u}_m$ with $m\in\{1,\ldots, M\}$. The received power at the UE position is given as
\begin{multline} 
P_{\text{UE}}={P_{\text{BS}}\frac{G_{\text{BS}} G_{\text{UE}}(d_\text{y} d_\text{z})^2}{16\pi ^{2}}}\\
\times\left | \sum_{m=1}^{M}\Gamma_{m}{\frac{\sqrt{F_{m}^{\text{c}}}e^{-j2\pi (|\Vec{a}-\Vec{u}_m|+|\Vec{b}-\Vec{u}_m|)/\lambda}}{|\Vec{a}-\Vec{u}_m||\Vec{b}-\Vec{u}_m|}}\right |^{^{2}}.
\label{eq:ReceivedPower}
\end{multline}
where $P_{\text{BS}}$, $G_{\text{BS}}$, $G_{\text{UE}}$, $d_\text{y}$, and $d_\text{z}$ are the transmit power of the BS, the BS and UE antenna gain as well as the effective RIS element dimension in $y$ and $z$ direction, respectively \cite{Tang21, Tang22, Radpour24}. The complex reflection coefficient of each RIS element $m$ is denoted by $\Gamma_{m}$. The wavelength $\lambda=c_0/f$, where $f$ denotes the center frequency and $c_0$ the speed of light. The combined antenna pattern of the BS antenna, the RIS element $m$ for receive and transmit operation, as well as the UE antenna is described by
$F^\text{c}_m$, please see \cite[(2)]{Radpour24} for details.
 
\section{RIS Design}
\label{sec:RIS Design}
\begin{figure}
	\centering
	\includegraphics[width=.7\columnwidth]{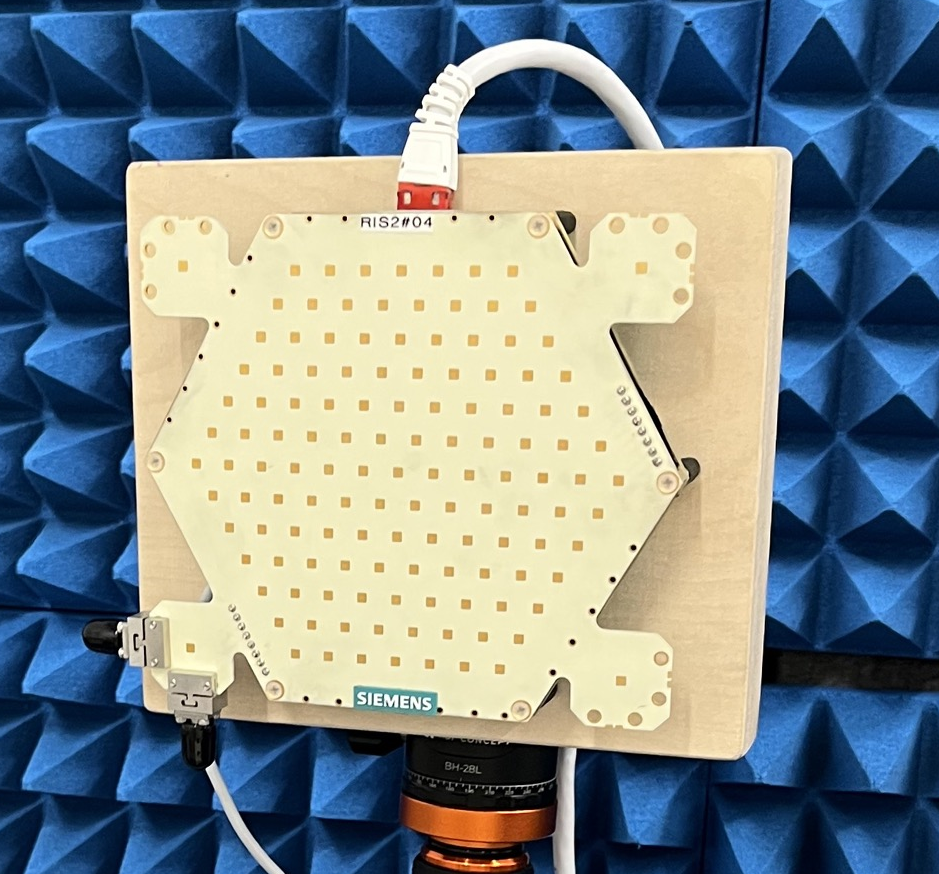}
	\caption{Closeup of the RIS printed circuit board with 127 RIS elements.} 
	\label{fig:RIS_closeup}
\vspace{-4mm}
\end{figure}

The RIS PCB, depicted in Fig. \ref{fig:RIS_closeup}, has been improved compared to the version presented in \cite{Radpour24} where 37 elements were used. In the current version, the RIS consists of $M=127$ elements arranged in five hexagonal rings with an additional element positioned at the center. The RIS operates at a center frequency of 23.8\,GHz. Table \ref{tab:ActiveRIS} provides a summary of the RIS parameters and the measurement setup. The origin of the coordinate system is considered to be the RIS center.
\begin{table}
\begin{center}
\caption{Active RIS Parameters and Measurement Setup.}
\label{tab:ActiveRIS}
\begin{tabular}{ll} 
\toprule
Parameter	      	&  Definition\\
\midrule
$f = 23.8$\,GHz		& 	 center frequency \\
$M = 127$    & number of RIS elements\\
$d_z$, $d_y = 6.6$\,mm	& 	 effective RIS element size \\
$d = 8.7$\,mm & smallest RIS element distance\\
\midrule
$P_\text{BS}=10$\,dBm		& 	 BS transmit power\\
$G_\text{BS} = 19$\,dB		&  BS horn antenna gain\\
$G_\text{UE} = 3.2$\,dB		&   UE vert. pol. monopole antenna gain\\
$|\Vec{a}|= 1.86$\,m,\,$|\Vec{b}|= 1.4$\,m  & distance RIS-BS and RIS-UE\\
$\tilde{\Vec{a}}    =( 1.86\,\text{m}, -36^\circ, 0^\circ)$ & BS location\\
$\tilde{\Vec{b}}_{P_1}=(1.4\,\text{m}, 40^\circ, -16^\circ)$ &  UE location at point $P_1$\\
$\tilde{\Vec{b}}_{P_2}=( 1.4\,\text{m}, 10^\circ, -16^\circ)$ & UE location at point $P_2$\\
\bottomrule
\end{tabular}
%\vskip-20pt
\end{center}
\end{table}
The matching network in \cite[Fig. 4]{Radpour24} has been modified such, that we can use the following reflection coefficient alphabet for the reflective mode 
$$\mathcal{A}_\text{R}=\{(0.3, \angle -15^\circ) , (0.3, \angle 165^\circ) \}.$$ Hence, the phase difference between both states is now $180 ^\circ$, allowing to cancel the specular reflection, cf. \cite[Sec. III]{Radpour24}. In active mode the RIS hardware achieves
$$\mathcal{A}_\text{A}=\{(1.25, \angle 0^\circ) , ( 0, \angle 0^\circ) \}.$$ Hence, in active mode the reflected signal of an element is $1.25/0.3\simeq4.2\simeq6.2\,\text{dB}$ stronger than in reflective mode.

The reflection coefficients for each of the RIS elements are chosen to maximize the power at a desired UE position through a coherent superposition of all RIS elements contribution, cf. \cite[Algorithm 1]{Radpour24}.

\section{RIS Power Pattern Evaluation}
\label{sec:RISpowerpattern}
\begin{figure}
	\centering
	\includegraphics[width=.9\columnwidth]{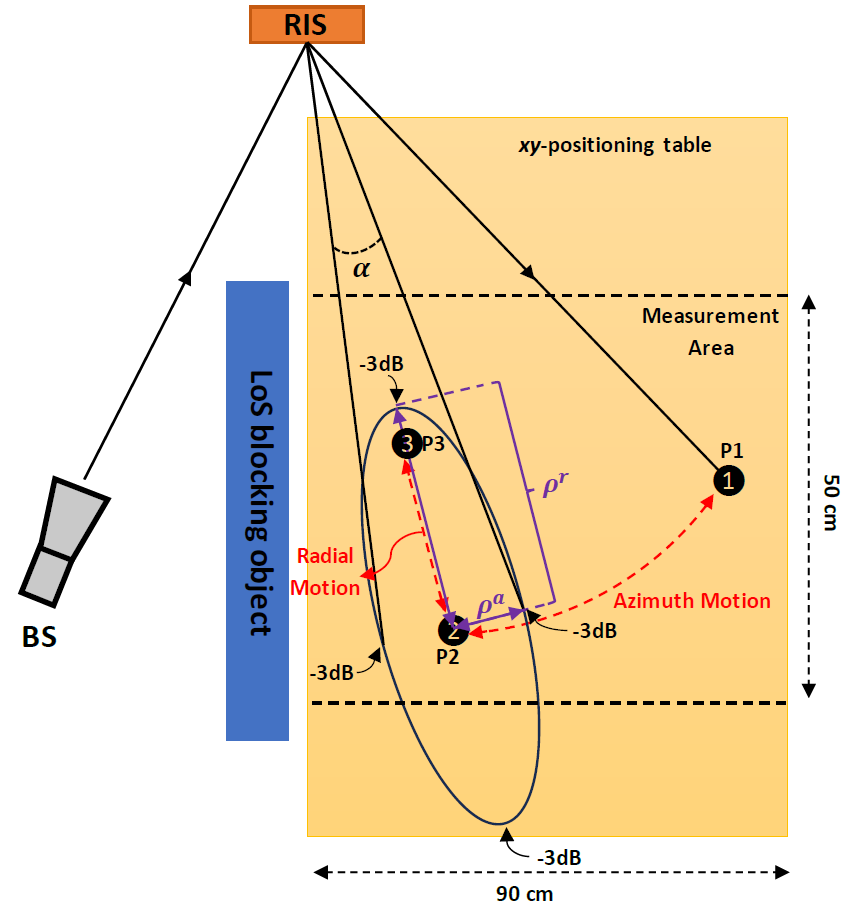}
	\caption{RIS testbed top view} 
	\label{fig:RIStestbed}
\vspace{-4mm}
\end{figure}

\begin{figure*}
	\centering
	\subfloat[RIS measurement testbed with LoS blockage.]{\includegraphics[width=0.35\paperwidth]{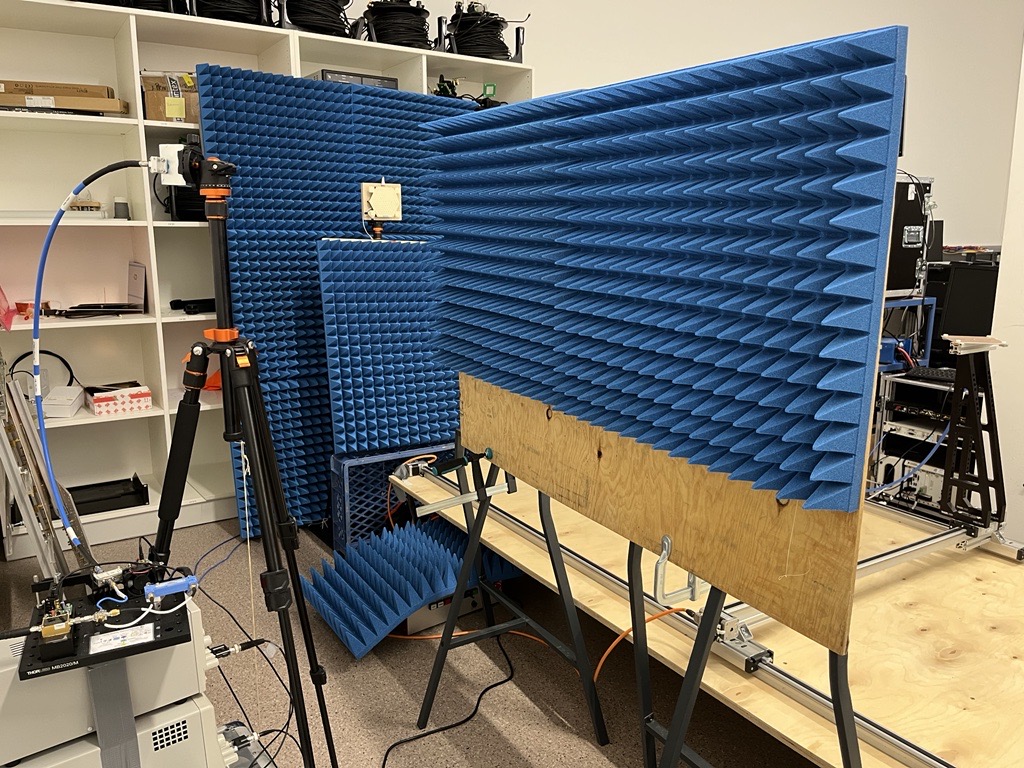}\label{fig:wBlockage}}
	\subfloat[RIS Measurement testbed without blockage.] {\includegraphics[width=0.35\paperwidth]{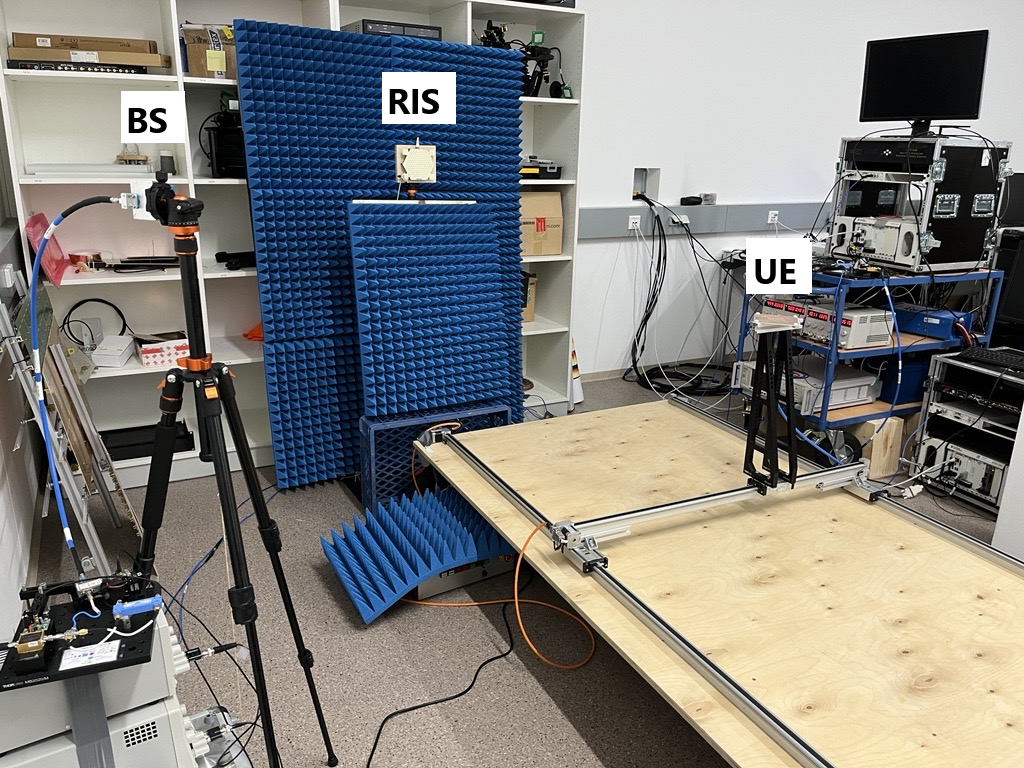}\label{fig:woBlockage}}
    \caption{Measurement testbed with (a) and without (b) the LoS blockage object.}
	\label{fig:RealMeasurementSetup}
	\vskip-20pt
%	\vspace{-4mm}
\end{figure*}

To evaluate the performance of the RIS in an industrial scenario, we evaluate the received power pattern $P_\text{UE}$, with an $xy$-positioning table measuring the received power $P_\text{UE}(\tilde{\Vec{b}})$. We configure the RIS elements to maintain the received signal power at defined coordinates $\tilde{\Vec{b}}\in \left \{ \tilde{\Vec{b}}_{P_1}, \tilde{\Vec{b}}_{P_2}\right \}$. A schematic representation of the measurement setup is shown in Fig. \ref{fig:RIStestbed}. 
The transmitter (TX) horn antenna, mounted on a tripod is pointing to the center of the RIS that focuses the signal on the desired UE position. For this measurement setup, the UE is a vertically polarized omnidirectional antenna, that is mounted on the $xy$-positioning table and can move to defined positions. An absorber blocks the direct LoS between TX and RX, simulating a NLoS industrial scenario. A picture of the actual measurement setup in the laboratory is shown in Fig. \ref{fig:RealMeasurementSetup}. 

While the TX location is fixed, the UE is moving on the $xy$-positioning table with the sampling steps of 2\,\textup{cm} in the measurement area of $0.92\leq x\leq 1.52$\,m and $0.02\leq y\leq 0.92$\,m. This means that we have 1380 sampling points in the measurement area. For the numerical simulation with \eqref{eq:ReceivedPower}, we modeled the schematic measurement setup and calculated the received power at the same positions as the measurement.

We consider two target points $P_1$ and $P_2$, with coordinates $\tilde{\Vec{b}}_{P_1}=(1.4\,\text{m},40^\circ,-16^\circ)$ and  
$\tilde{\Vec{b}}_{P_2}=( 1.4\,\text{m}, 10^\circ, -16^\circ)$, respectively. We chose the points $P_1$ and $P_2$ such that they have the same distance from the RIS and the only difference is their azimuth angle. This is because we want to know often we need to reconfigure the RIS to maintain $P_\text{UE}$ above a desired minimum power threshold. In Section \ref{sec:SimMeasResult} we will present and compare the simulation and measurement results for the RIS, focusing the signal on the points $P_1$ and $P_2$. Furthermore, we show the results for the RIS working in reflective and active mode.

\subsection{Numerical Power Pattern Simulation}
We simulate the power pattern at a given coordinate $\Vec{b}$ using the formula in \eqref{eq:ReceivedPower}. We obtain the configuration of the RIS elements using the optimization algorithm presented in \cite[Algorithm 1]{Radpour24}. For the simulation, we evaluate the receive power with a resolution of 2\,cm in $x$ and $y$ directions.

\subsection{Empirical Power Pattern Measurement}
We perform a measurement of the impulse response (IR) $h_{x,y}[n]$ vs. $x$ and $y$ coordinates on the $xy$-table. We perform the measurement such that the mobile UE moves from one point to the next adjacent point in the $y$-direction. Then we insert a pause time of $\Delta t=200\, \text{msec}$ to let the possible UE vibrations fade away after the movement. Next, we perform $Q$ repetitive measurements to record the channel sounder data. Then we take the average of the recorded IRs and calculate the received power as
\begin{equation}
    P_\text{UE} (x,y)=\frac{P_\text{BS}}{Q}\left\vert \sum_{n=n_1}^{n_2} \sum_{q=1}^{Q} h_{x,y}^{q}[n]\right\vert^2.
\end{equation}
to obtain the RIS power pattern, we set $Q=50$ from which we calculate the average of the 50 recorded measurements, $n_1$ and $n_2$ define the support of the impulse response above the noise floor to suppress measurement noise as much as possible. The impulse response measurements in this paper have a support in the interval $n_1\leq n \leq n_2$ with $n_1= 7$ and $n_2 = 13$. 

\section{Simulation and Measurement Result Analysis} 
\label{sec:SimMeasResult}
Initially, we carefully calibrated the RIS orientation in 3D space such that the RIS surface norm in the center matches the $x$-axis in a positive direction. This is important to achieve a good alignment of the received power pattern between measurement and numerical simulation. Then we performed the experiment in four steps:
\subsection{Validating non-Line-of-Sight Scenario}
\label{sec:woRIS}
\begin{figure*}
	\centering
	\subfloat[Measurement result with the RIS removed from the measurement setup.]{\includegraphics[width=0.4\paperwidth]{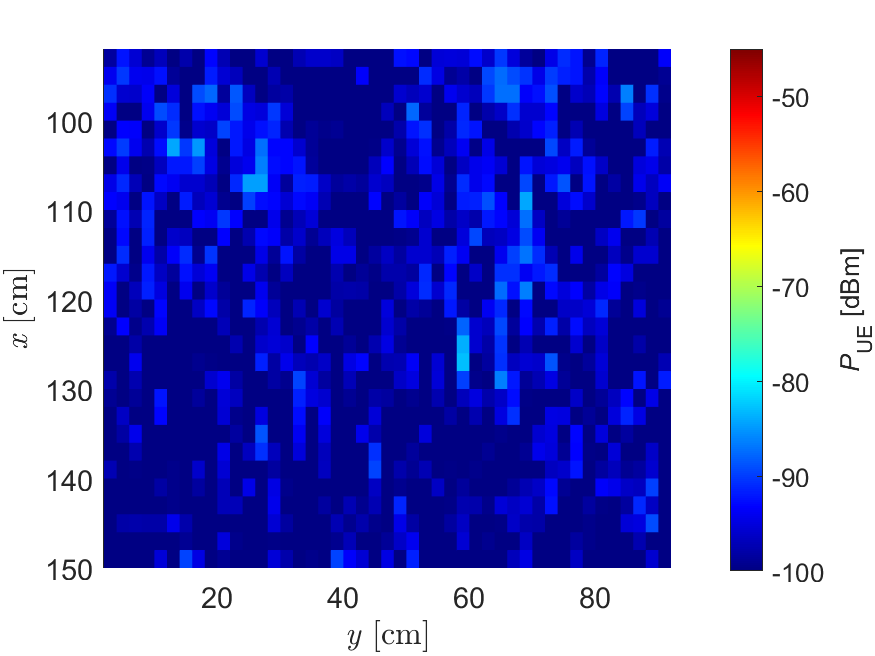}\label{fig:woRIS}}
	\subfloat[Measurement result with the RIS switched off.]{\includegraphics[width=0.4\paperwidth]{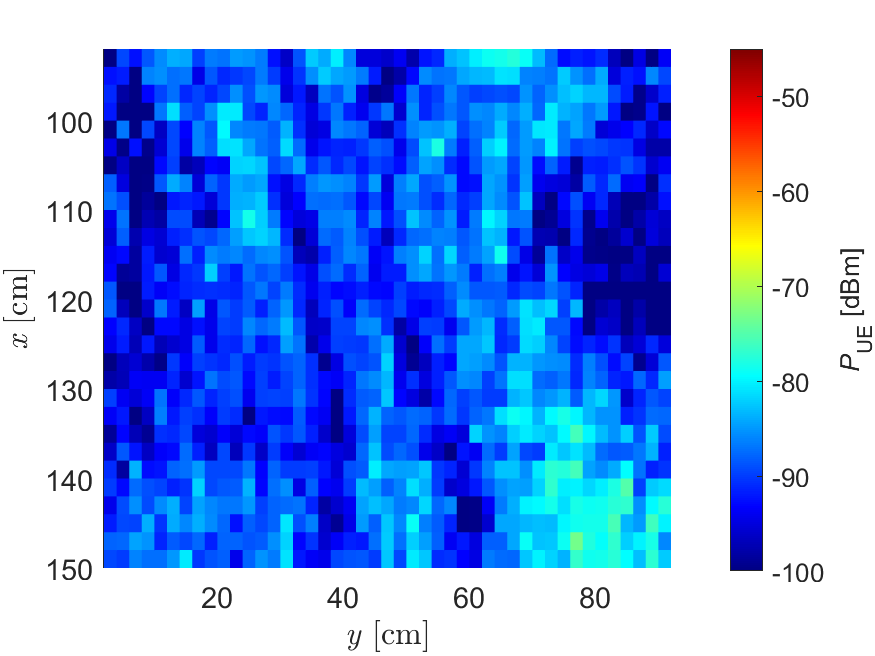}\label{fig:SwitchOFFRIS}}
    \caption{Received power $P_\text{UE}(x, y)$ over measurement area of $xy$-positioning table. We illustrate the empirical measurement data for (a) without RIS and (b) RIS switched off}
	%\label{fig:SwitchOFFRIS}
		\vspace{-4mm}
\end{figure*}

To evaluate the received power at the UE without RIS in a NLoS scenario, we physically removed the RIS from the testbed shown in Fig. \ref{fig:wBlockage}. In Fig. \ref{fig:woRIS} we observe that the received power at most $(x,y)$ positions is as low as $P_\text{UE}^{\text{meas}}=-100\,$dBm. This is because the absorber blocks the majority of the transmitted power. Hence, we will use a RIS to circumvent the blocking obstacle and obtain a higher received power.
The measurement noise power is well inline with the theoretical value 
\begin{equation}
P_\text{UE}^{\text{meas}} = -100 \, \text{dBm} \approx 10 \log10\left(\frac{k T B}{ 1 \text{mW} \cdot Q}\right) + N_\text{F}
\end{equation}
with the Boltzmann constant $k=1.38\cdot 10^{-23} \frac{\text{m}^2 \text{kg}}{\text{s}^2 \text{K}}$, temperature $T=293\,\text{K}$, bandwidth $B=155\,$MHz and a combined noise figure of the TX and RX amplifier of $N_\text{F}=9\,$dB.

\subsection{Validating Limited Reflection of RIS when Switched Off}
\label{sec:Swithched_OFF}
In order to assess the effect of the RIS while it is switched off, we mounted the RIS but turned off its power supply. Figure \ref{fig:SwitchOFFRIS} shows that due to the structural reflectivity of the RIS elements, we get a small amount of received power within the measurement area. According to Fig. \ref{fig:SwitchOFFRIS}, the maximum received power is $P_\text{UE}^{\text{meas}}=-80\,$dBm around the position of the specular reflection. 

\subsection{RIS Operating in Reflective Mode}
\label{sec:ReflectiveModeResult}
\begin{figure*}
	\centering
	\subfloat[Sim. result for reflective RIS, focusing signal on point $P_1$.]{\includegraphics[width=0.4\paperwidth]{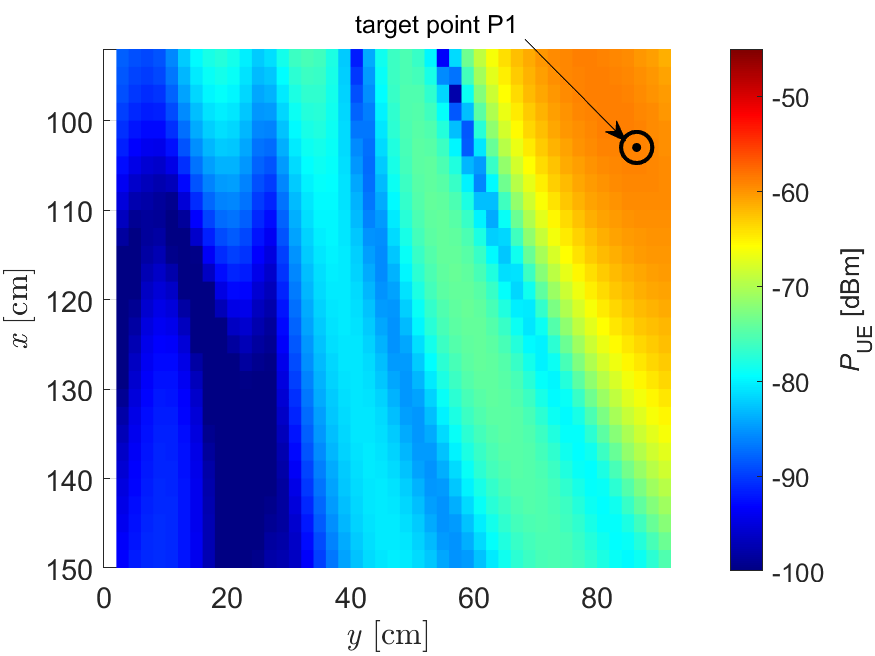}\label{fig:RefSimP1}}
	\subfloat[Meas. result for reflective RIS, focusing signal on point $P_1$.]
 {\includegraphics[width=0.4\paperwidth]{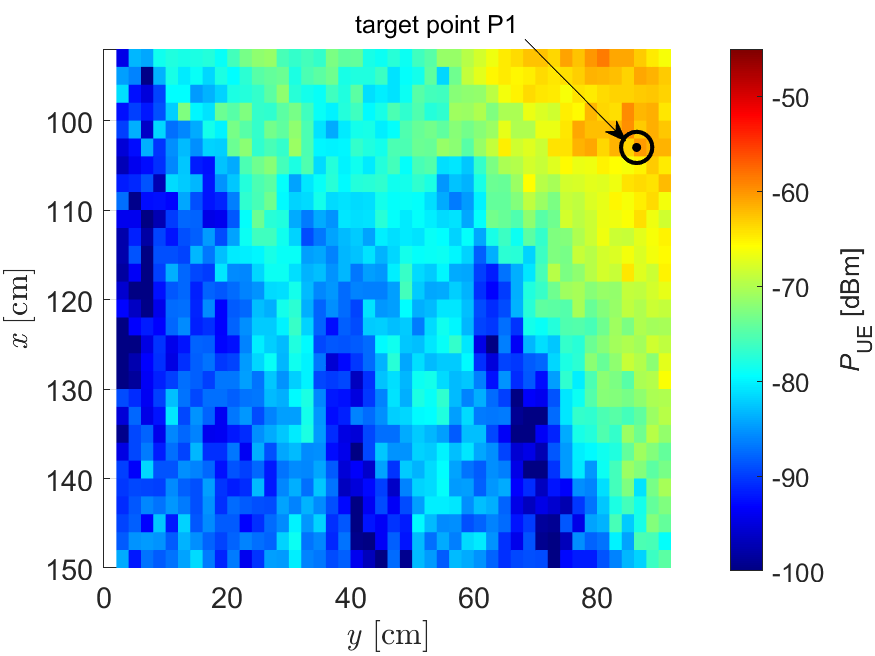} \label{fig:RefMeasP1}}\\[-2ex]
 	\subfloat[Sim. result for reflective RIS, focusing signal on point $P_2$.]{\includegraphics[width=0.4\paperwidth]{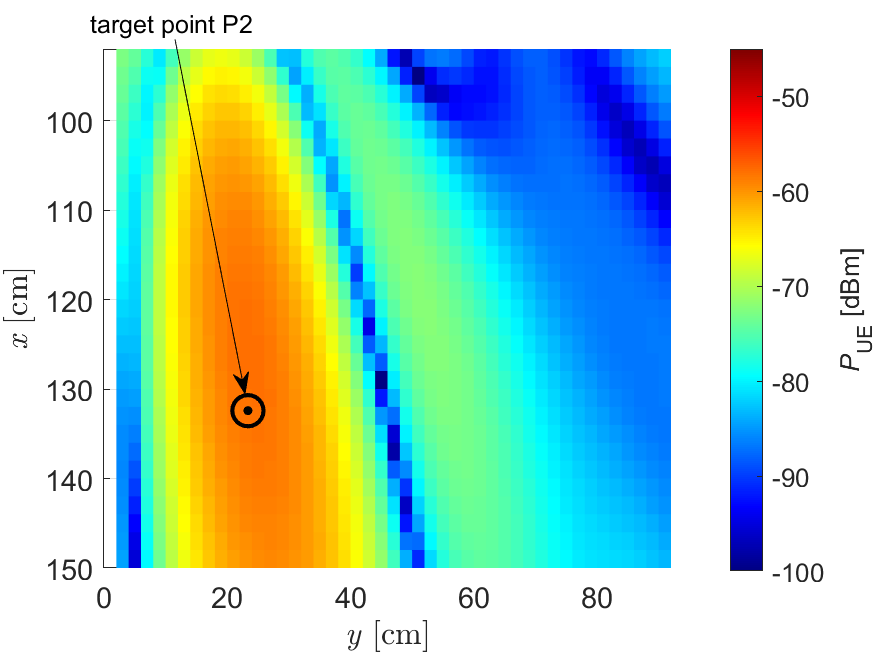}\label{fig:RefSimP4}}
	\subfloat[Meas. result for reflective RIS, focusing signal on point $P_2$.]
 {\includegraphics[width=0.4\paperwidth]{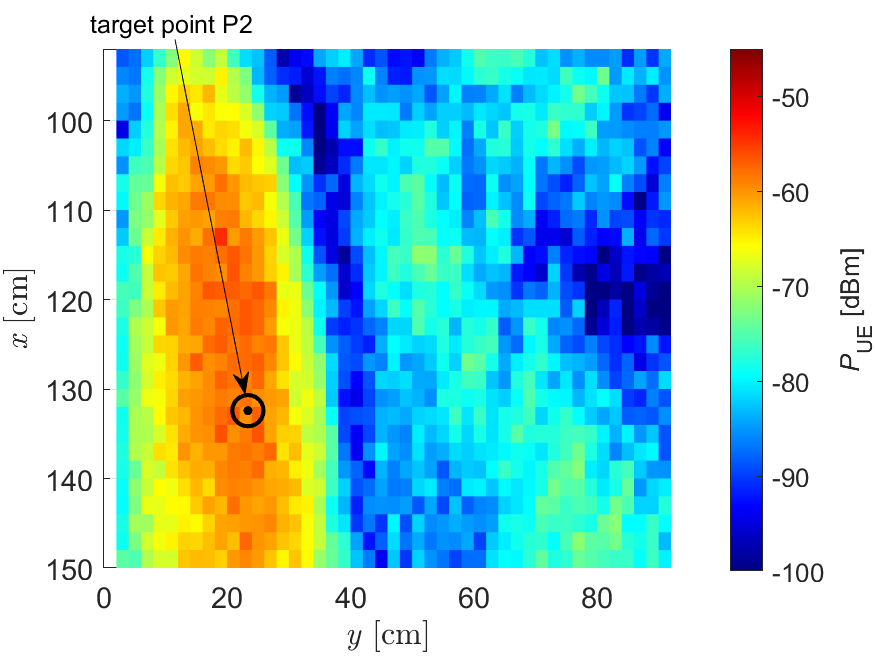}\label{fig:RefMeasP4}}
    \caption{Received power $P_\text{UE}(x, y)$ over measurement area of $xy$-positioning table. We compare the numerical simulation results with empirical measurement data for the RIS in reflective mode. The BS is located at $\tilde{\Vec{a}}= ( 1.86\,\text{m},-36^\circ, 0^\circ)$ and the signal shall be focused on the UE positions $\tilde{\Vec{b}}_{P_1}=(1.4\,\text{m},40^\circ,-16^\circ)$ and $\tilde{\Vec{b}}_{P_2}=(1.4\,\text{m},10^\circ,-16^\circ)$.}
	\label{fig:RefResult}
		\vspace{-4mm}
\end{figure*}

In reflective mode, the RIS elements are set to have a phase shift of either $-15^\circ$ or $+165^\circ$. This ability to have a phase shift difference of $180^\circ$ improves the RIS performance in the reflective mode compared to the results in \cite{Radpour24}, allowing to suppress the specular reflection. In Figs. \ref{fig:RefSimP1} and \ref{fig:RefMeasP1} both the simulation and the measurement result show that the RIS can focus the radio signal on point $P_1$ and the received power for simulation and measurement are $P_\text{UE}^{\text{sim}}=-57\,$dBm and $P_\text{UE}^{\text{meas}}=-60\,$dBm, respectively. In Figs. \ref{fig:RefSimP4} and \ref{fig:RefMeasP4} the RIS is focusing the radio signal on the target point $P_2$. The mobile UE in simulation and measurement has a receive power of $P_\text{UE}^{\text{sim}}=-56\,$dBm and $P_\text{UE}^{\text{meas}}=-57\,$dBm, respectively. These results show that the reflective RIS can provide an SNR of more than 40 dB at the intended target point. The comparison between the simulation and measurement results illustrates not only a high quantitative match but also the power levels of the simulation and measurement match with high accuracy.
\subsection{RIS Operating in Active Mode}
\label{sec:ActiveModeResult}
\begin{figure*}
	\centering
	\subfloat[Sim. result for active RIS, focusing radio signal on point $P_1$.]{\includegraphics[width=0.4\paperwidth]{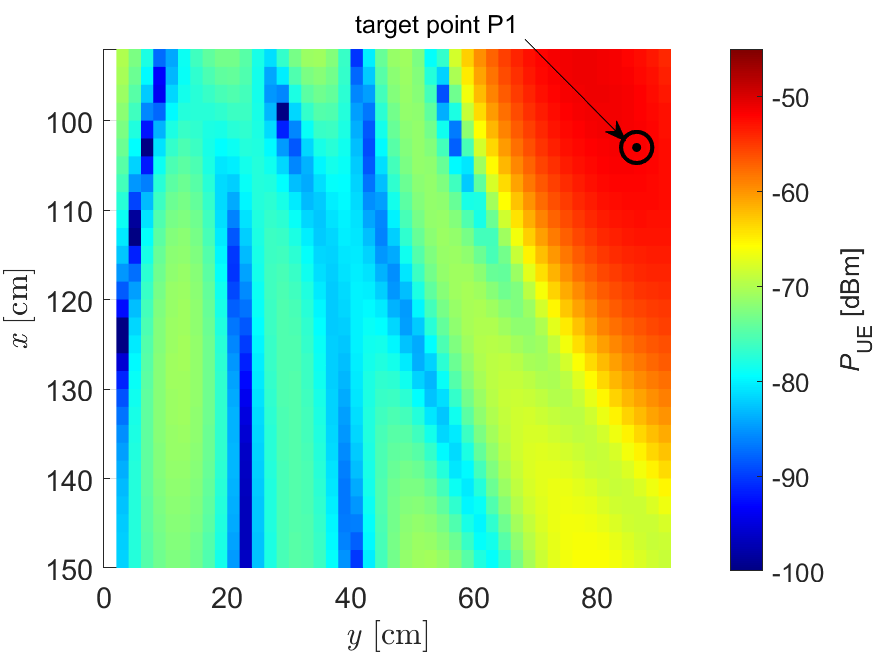}\label{fig:ActiveSimP1}}
	\subfloat[Meas. result for active RIS, focusing radio signal on point $P_1$.]
 {\includegraphics[width=0.4\paperwidth]{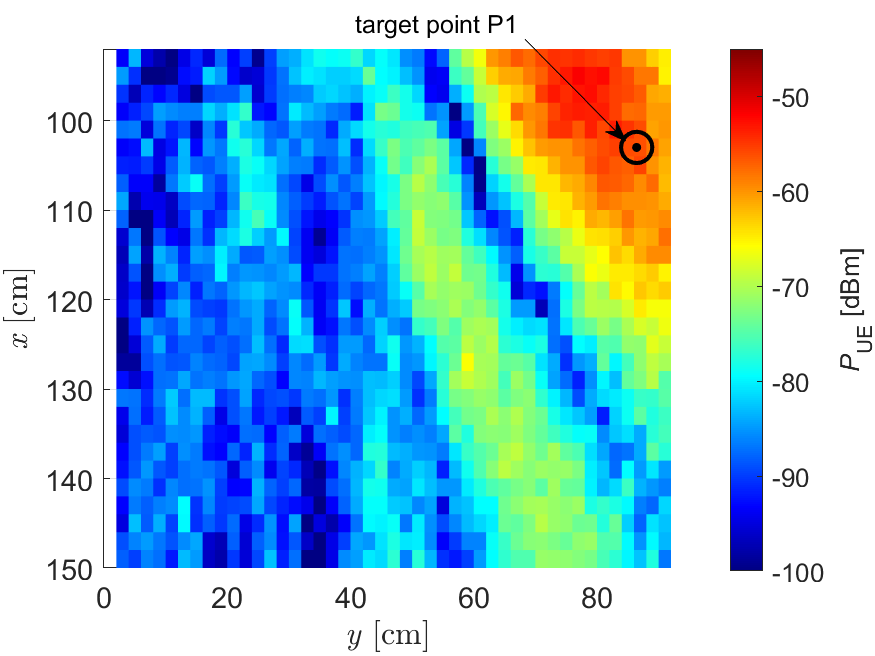}\label{fig:ActiveMeasP1}}\\[-2ex]
 	\subfloat[Sim. result for active RIS, focusing radio signal on point $P_2$.]{\includegraphics[width=0.4\paperwidth]{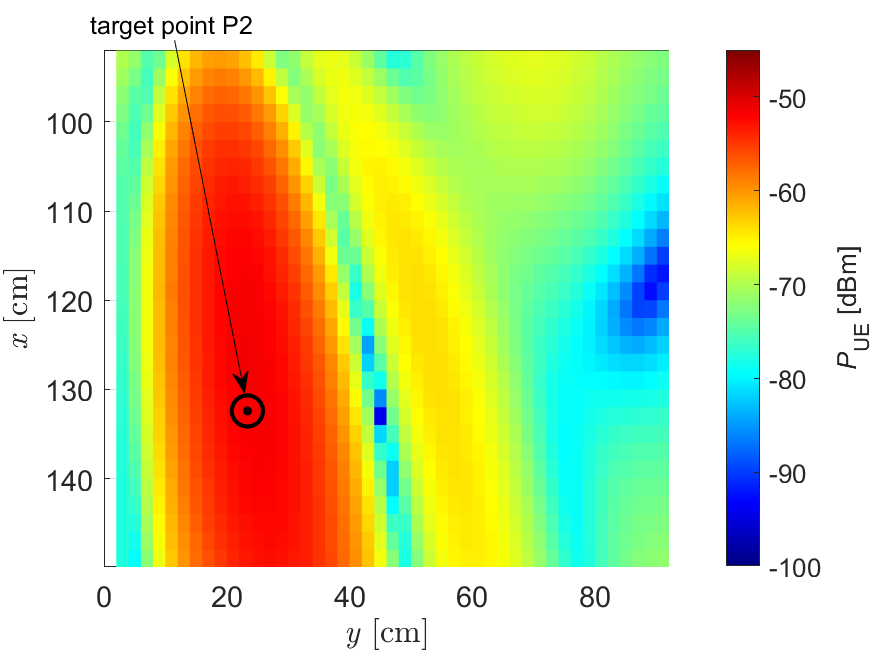}\label{fig:ActiveSimP4}}
	\subfloat[Meas. result for active RIS, focusing radio signal on point $P_2$.]
 {\includegraphics[width=0.4\paperwidth]{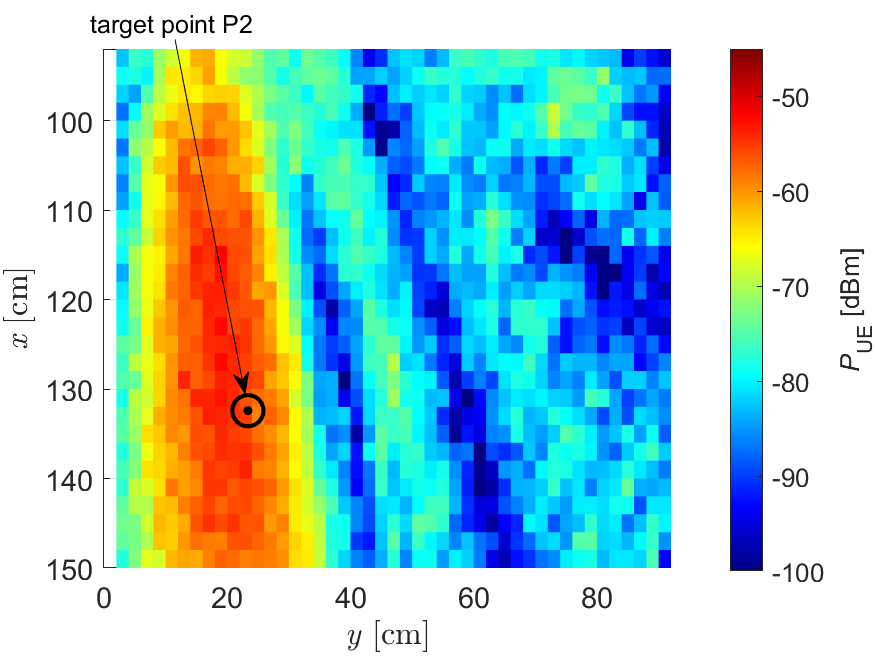}\label{fig:ActiveMeasP4}}
    \caption{Received power $P_\text{UE}(x, y)$ over measurement area of $xy$-positioning table. We compare the numerical simulation results and empirical measurement data for the RIS in active mode. The BS is located at $\tilde{\Vec{a}}= ( 1.86\,\text{m},-36^\circ, 0^\circ)$ and the signal is focused on the UE position $\tilde{\Vec{b}}_{P_1}=(1.4\,\text{m},40^\circ,-16^\circ)$ and $\tilde{\Vec{b}}_{P_2}=(1.4\,\text{m},10^\circ,-16^\circ)$.}
	\label{fig:ActivefResultP}
		\vspace{-4mm}
\end{figure*}

In active mode, the RIS elements are either turned off or they amplify the received signal with a gain of 1.25. Figures \ref{fig:ActiveSimP1} and \ref{fig:ActiveMeasP1} show the power patterns of the RIS working in active mode while focusing the radio signal on the target point $P_1$. The power level at the mobile UE in simulation and measurement is $P_\text{UE}^{\text{sim}}=-52\,$dBm and $P_\text{UE}^{\text{meas}}=-56\,$dBm, respectively. Finally, Figs. \ref{fig:ActiveSimP4} and \ref{fig:ActiveMeasP4} show the results while the RIS is focusing the radio signal on point $P_2$ with a received power of  $P_\text{UE}^{\text{sim}}=-52\,$dBm and  $P_\text{UE}^{\text{meas}}=-53\,$dBm in simulation and measurement, respectively.  Therefore, using an active RIS in this NLoS scenario provides an SNR level of at least 44 dB at the desired point. Finally, a comparison between the received power levels in the reflective and active results indicates that the active RIS outperforms the reflective RIS by a gain of at least 4\,dB.

\section{Update Requirement for RIS Configuration in Industrial Automation and Control Scenario}

For a RIS-assisted wireless communication system with a mobile UE in an industrial environment, we need to know how often the RIS needs to be reconfigured. This is crucial for the mobile UE to maintain received power above a certain threshold. We calculate the coverage for a mobile UE, moving radially (motion along the RIS) or moving in azimuth direction (motion by changing azimuth angle but keeping the same distance from the RIS); please see Fig. \ref{fig:RIStestbed} for more details. Here, we set the threshold to be the half power beamwidth (HPBW) and calculate the RIS focusing area as
 
\begin{equation}
    \rho^{r}=\frac{1}{2}\left(\frac{h_{\textup{RIS}}-h_{\textup{UE}}}{\tan(\theta _{\textup{UE}}-\beta/2)}-\frac{h_{\textup{RIS}}-h_{\textup{UE}}}{\tan(\theta _{\textup{UE}}+\beta/2)}\right)
    \label{eq:2}
\end{equation}
\begin{equation}
    \rho^{a}=|\Vec{b}|\tan(\alpha/2)
    \label{eq:3}
\end{equation}
where $\rho^{\text{a}}$ and $ \rho^{\text{r}}$ are corresponding to the azimuth and radial motion and define the minor and major axis of an ellipse that specifies the RIS HPBW focusing area on the $xy$-positioning table, $h_{\textup{RIS}}$ and $h_{\textup{UE}}$ are the RIS and UE heights, $\theta _{\textup{UE}}$ is the UE elevation angle, and $\beta$ and $\alpha$ are the vertical and horizontal HPBW angles, respectfully, (please see Fig.\ref{fig:RIStestbed}). 

Considering the RIS center as the origin of the coordinate system, for our setup we have $(h_{\textup{RIS}},h_{\textup{UE}})=(0\,\textup{cm},-39\,\textup{cm})$ and $\theta _{\textup{UE}}=-16^\circ$. Applying \eqref{eq:2} and \eqref{eq:3} to the results shows that $(\alpha, \beta)\simeq (8^{\circ},7^{\circ})$ and the dimension of the focus area in the radial and azimuth directions are $\rho^{\text{a}}\simeq9\,\textup{cm}$ and $\rho^{\text{r}}\simeq 40\, \textup{cm}$, i.e., the size of the focusing area is smaller in the azimuth direction than in the radial direction. Therefore, for a mobile UE that moves in the azimuth direction, the RIS requires more configuration updates than for radial motion. It is important to mention that the HPBW angles are variable and depend on the RIS configuration. For a mobile robot located at point $P_2$ moving in an azimuth direction with a speed of $v=1\,\textup{m/sec}$ towards the point $P_1$, we need to update the RIS setting every $\rho^{\text{a}}/v = 90\,\textup{msec}$ to be in the RIS focus area. However, if the robot moves with the same speed towards the point $P_3$, we need to update the RIS setting every $\rho^{\text{r}}/v= 400\,\textup{msec}$. 

\section{Conclusion}
We tested the capability of a RIS proof-of-concept (PoC) that allows establishing a reliable link in mmWave frequencies for a mobile UE in an indoor NLoS scenario. We measured the received power for a mobile UE moving on an $xy$-positioning table. We evaluated the received power at the UE for the RIS working in reflective or active mode. We have analyzed the control algorithm requirements for the RIS to serve a mobile UE in a NLoS industrial environment, ensuring that the received power at the UE location exceeded a specified threshold. Our analysis of the received power pattern, based on both simulations and measurements, revealed that the RIS update rate must be four times higher for a mobile UE moving in the azimuth direction compared to radial movement. The results from both simulations and measurements demonstrated that RIS effectively focused the radio signal on the desired targets and that the active RIS provides 4\,dB higher received power compared to the optimized reflective RIS.

\section*{Acknowledgment}
This work is funded through the Vienna Business Agency in the project RISING and by the Principal Scientist grant at the AIT Austrian Institute of Technology within the project DEDICATE.

%\bibliographystyle{IEEEtran}

%\bibliography{IEEEabrv, DissLib}   

\end{document}